# Measuring and simulating latency in interactive remote rendering systems

Richard Cloete, Nick Holliman Member IEEE Computer Society

**Abstract**— Background—The computationally intensive task of real-time rendering can be offloaded to remote cloud systems. However, due to network latency, interactive remote rendering (IRR) introduces the challenge of interaction latency (IL), which is the time between an action and response to that action. Objectives—to model sources of latency, measure it in a real-world network and to use this understanding to simulate latency so that we have a controlled platform for experimental work in latency management. Method— we present a seven-parameter model of latency for a typical IRR system; we describe new, minimally intrusive software methods for measuring latency in a 3D graphics environment and create a novel latency simulator tool in software. Results— We demonstrate our latency simulator is comparable to real-world behavior and confirm that real-world latency exceeds the interactive limit of 70ms over long distance connections. We also find that current approaches to measuring IL are not general enough for most situations and therefore propose a novel general-purpose solution. Conclusion—to ameliorate latency in IRR systems we need controllable simulation tools for experimentation. In addition to a new measurement technique, we propose a new approach that will be of interest to IRR researchers and developers when designing IL compensation techniques.

——————————— ◆ ———————————

## 1 INTRODUCTION

Measuring Interaction Latency (IL) in interactive Cloud-based visualization systems, which we will refer to as an Interactive Remote Rendering (IRR) systems [1], is a crucial yet challenging task. Latency is typically not controllable, making it difficult to study compensation techniques, their effects on user performance and their impact on user experience. Additionally, current latency measurement techniques are not general-purpose enough, which further complicates system performance benchmarking. We aim to introduce a new method for measuring and simulating latency in IRR systems that fulfils these goals.

Today, 3D graphics are ubiquitous in computing, found everywhere from calculators and smart-watches, to smart-glasses. Rendering graphics locally can facilitate highly responsive user experiences, but often requires significant compute resources which are not always available to the user.

Remote rendering, where rendering is performed on powerful remote servers, has been employed by the film and animation industry for three decades [2]. Specifically, this type of system does not require user interaction and therefore data can be processed and rendered, without having to respond in real-time to user input: the results (images) can be saved to disk and used at a later time. These non-interactive remote rendering systems are known as render farms [3].

To achieve interactive performance from remote rendering systems we need to ameliorate interaction latency. Therefore, in this paper we investigate in detail the sources, measurement and simulation of latency, and make the following novel contributions:
- A model for interaction latency.
- A method for simulating interaction latency.
- A novel framework for measuring interaction latency using a software-based observer approach, as well as for testing and validation.

———————————

- R. Cloete is with the School of Computing, Newcastle University, Newcastle, NE4 5TG, UK. E-mail: r.cloete1@newcastle,ac.uk.
- N.S. Holliman is with the School of Computing, Newcastle University, NE4 5TG, E-mail: nick.holliman@newcastle.ac.uk

## 2 WHAT IS AN INTERACTIVE REMOTE RENDERING SYSTEM?

Interactive Remote Rendering (IRR) systems are those which enable a user of a local thin-client device (laptop, tablet, smartphone, etc.) to interact with data which are hosted, processed and rendered remotely, on powerful servers in the cloud [1]. Such systems leverage the power of Cloud Computing to perform computationally intensive rendering tasks and thereby free thin client devices of potentially significant resource constraints, enabling users to explore and interact with massive data sets.

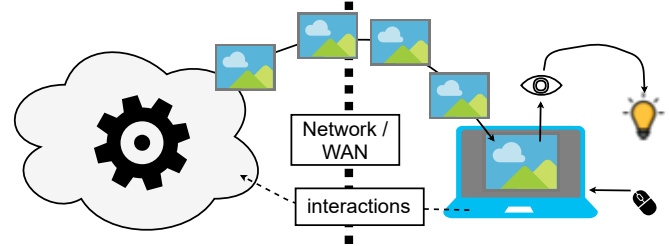

Figure 1. A simplistic IRR system architecture. Interactions such as mouse-clicks are performed on the thin client device. Those interaction are transmitted across a network to a powerful remote server which processes, renders and then transmits an image result back to the client where it is displayed to the user, leading to insights and idea generation.

In the last 10 years, IRR systems have attracted a lot of attention [4][5][6][7]. Rapid advances made in mobile device and communications (2G, 3G, 4G and soon-to-be 5G) technology, and the potential to make online video gaming and data visualization more accessible to consumers has undoubtedly contributed to this interest. However, despite the technological progress made over the past decade, mobile device capabilities are still very limited when compared with dedicated servers. For example, a Nvidia RTX 1080 provides teraflop performance [8], while the latest smartphone GPU (Qualcomm Adreno 630) provides gigaflop performance [9].

While IRR systems do offer significant advantages to local rendering, they introduce a new challenge: Interaction Latency (IL).

## 3 Defining Interaction Latency (IL)

Interaction Latency (IL) is the difference in time between the moment an interaction is registered by the client device (i.e. a laptop), and the point at which the corresponding frame result is displayed to the user.

Academic interest in IRR systems has resulted in researchers [10][11][12][13][14] identifying IL as a significant challenge, with studies showing that users are able to detect IL as low as 2.38ms [15] and that users will interpret an action as the cause of an event as long as the IL experienced is less than 70ms. However, when IL exceeds 160ms, users experience a disconnect between events and the actions that caused them [16], possibly leading the user to become confused.

In video games, users have been shown to play far less when experiencing IL in excess of 250ms, compared to those who experienced IL of 150ms [17]. Thus, as little as 100ms IL can be responsible for a loss of up to 75% user engagement. IL is also a concern for exploratory visual analysis. For example, in [11], Liu et al. show that an IL greater than 500ms results in decreased user activity, poorer data set coverage and reduces rates of observation, generalization and hypothesis formation.

A combination of factors impacts IL, with Network Latency (NL), being the most significant contributor. Unfortunately, NL is inherent in all network distributed applications and is dictated by various factors such as the physical distance between the client and server machines, and the type of material (i.e. copper vs fiber) used to transfer data across the network [18].

The most common approach to dealing with IL is to simply have servers located as near to their target users as possible. Unfortunately, this is not always possible as geographic, economic and ethical constraints may be prohibitive [19].

Minimizing IL is a key challenge for IRR system designers as paying close attention to it is critical for those who want to ensure that their products meet strict latency requirements, and that user engagement does not suffer due to poor responsiveness. It is therefore important that robust IL models and measurement tools are developed.

## 4 Modelling Interaction Latency

The main challenge in measuring IL is the need to identify the exact moment an interaction occurs, as well as when the corresponding frame update is visible to the user. This is difficult since in such systems, frame updates are typically the result of both user interaction and scene mechanics. For example, in a 3D racing game, an AI-controlled car may overtake a player independently of the user interactions; despite performing no action, frames independent of user input will continue to arrive from the server and be displayed on the client. Nevertheless, before IL can be measured, latency sources must be identified.

We identify five main sources of IL in typical IRR systems:

**Input Device Latency (IDL)** is the delay contributed by the input device, usually an external controller. This delay is the difference in time taken between an electronic signal (due to a button being pressed, for example) being generated on the hardware and having the input processed (memory updated in the OS event buffer). This delay may increase due to poor wireless connectivity, poorly charged batteries, etc.

**Client Latency (CL)** is the total delay experienced on the client device. Since $CL$ is the result of latency introduced both before interactions leave the client and after image results arrive, we identify that $CL$ is composed of two sub-latencies: $CL_1$ and $CL_2$. $CL_1$ is the delay from the point at which an interaction is received on the client, until that interaction exits the client. $CL_2$ is the delay between the client receiving a response from the server and processing (e.g. image decompression, client state updates) it, up to the point at which the pixels are pushed to the display.

**Network Latency (NL)** is the total delay caused by the transmission of a message between the client and the server. $NL$ occurs in both directions: from the client to the server ($NL_{up}$) and from the server back to the client ($NL_{down}$). The physical distance between the client and server is the primary source of $NL$. However, network routing, congestion and delivery (copper wire vs fibre optic, for example) all impact $NL$.

**Server Latency (SL)** is the total delay experienced on the server, from the point at which the interaction message is received, until a response message is generated and has left the server. $SL$ will always be present, however, poor rendering speeds, coding inefficiencies, image compression, low-end hardware, etc., will add to the total $SL$ and therefore negatively impact IL.

**Display Latency (DL)** exists in all monitors. It is the delay experienced between the display system receiving an input signal and the image being displayed on the screen. DL varies across display systems and is a combination of video input latency (time taken to perform upscaling and smoothing and double buffer switching (if any)) and pixel response time (the time taken for a physical pixel to update).

A general IRR pipeline model including the above latency sources is illustrated in Figure 2.

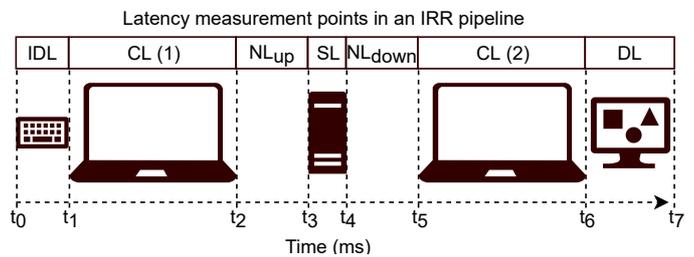

Figure 2. Interactive remote rendering latency model with measurement points. Interactions flow from left to right: input device to client to server, back to client and to the display. The sources of latency modelled are: Input Device ($IDL$), Client ($CL$), Network ($NL$) and Display ($DL$).

In Figure 2, an interaction occurs at $t_0$. At $t_1$ the interaction event is raised by the input device (such as a keyboard). The interaction signal is processed by the client application where at $t_2$, its transfer across a network is started. At $t_3$, the server receives the interaction message from the client and begins to process it. After a period of time, the processed and rendered results from the server are transmitted across the network; this begins at $t_4$ and ends at $t_5$, where it arrives on the client device. The client software then processes (decodes, verifies, warps, etc.) the results received from the server and at $t_6$, initiates the display process and passes the pixels to the monitor. Finally, at $t_7$, the pixels are presented on the display to the user. From this information, we can compute IL with $IL = t_7 - t_0$ or with:

$$IL = IDL + CL_1 + NL_{up} + SL + NL_{down} + CL_2 + DL \quad (1)$$

Interestingly, we can also calculate latency for various components. For example, we can determine $SL$ with:

$$SL = IL - (IDL + CL + NL + DL) \quad (2)$$

or with the following if DL is not known:

$$SL = (t_5 - t_0 - .5 * NL) - (t_2 - t_0 + .5 * NL) \quad (3)$$

However, we can still estimate $DL$ with:

$$DL = IL - (IDL + CL + NL + SL) \quad (4)$$

Unfortunately, manufacturers do not publish $DL$ information and instead, only supply response times. Nevertheless, we can estimate the maximum $DL$ from the monitor update frequency as a single frame time. NVIDIA's G-SYNC [20] technology aims to synch the GPU with the monitor refresh rate and will therefore significantly reduce $DL$.

## 5 MEASURING INTERACTION LATENCY

We identify that techniques for measuring IL fall into three primary categories: Integrated, Observer and Hardware.

### 5.2 Observer approaches

Observer approaches, such as one described by Chen in [21], involves writing software that "hooks" into the IRR system in order to take measurements. In their approach, the authors attempt to measure the IL of a cloud gaming system, which consists of a client application and a remote server. To achieve this, Chen et al. selected a game with a built-in menu screen which is activated by pressing the escape (ESC) key. To avoid the need for human input, the authors simulate the press of the ESC key with the expectation that after some period of time, a result from the server will arrive on the client and the client application will publish an on-screen update, thereby displaying the menu.

When the ESC key is simulated, a time measurement ($t_1$) is taken. To measure the moment the response from the server arrives ($t_5$), the authors hook into the recvfrom() function, which is called when the client attempts to retrieve data from the socket. The authors then measure the time between data arriving on the client and the frame being presented on-screen. To do this, a library called Detours is used to hook into the underlying graphics API. Specifically, the authors intercept the `IDirect3dDevice9::EndScene()` function, which is a function from the Direct3D library that is called just before graphics are about to be presented on-screen, and measure $t_6$. For each frame that arrives on the client, the colors of a chosen set of pixels are inspected. When a change in pixels is detected, they conclude that the screen has been updated and then record the time ($t_7$). The difference in time between user input and identifying the change in pixels yields an IL measurement.

One potential issue with the approach taken by Chen et. al is that their technique relies on there being an in-game menu. If the IRR system is not a games-based one, or if there is no screen update specific to a certain interaction, there is no way to determine whether or not the screen update is a result of an interaction or in-application mechanics. This issue may lead to incorrect measurements and potentially, make measurement impossible. Another issue is that, in order to obtain their measurements, the authors had to perform complicated hooking techniques which may result in instability of the IRR system and also requires expert knowledge to perform. A more general disadvantage of this approach is that hooking will change the system behavior, even if in a small way. Further, the authors did not compare their results with a system with known IL and therefore it is not clear how accurate or reasonable their results are.

### 5.3 Hardware approaches

Hardware approaches are those that aim to measure IL using hardware and/or external devices. For example, Steed in [23], uses a high-speed camera positioned in front of a monitor on which simulated frames of a 3D object will be displayed. A pendulum is then set between the monitor and the camera. The pendulum, fitted with a light-emitting diode (LED), is tracked and its position information is sent to and used by a rendering application to "drive" the simulated frame. Using the camera, the operator video-records the scene and analyzes the resulting footage. Finally, IL is estimated by "counting the number of frames between pendulum and the simulated image". Unfortunately, this technique requires the operator to set up equipment and have access to the required hardware (high-speed camera, tracker, LED, etc.), making it impractical for many users. Another potential problem is that counting frames is a manual process, performed by the operator, and is prone to human error. However, this was later addressed by Friston and Steed in [24], where they introduced an automatic frame counting algorithm.

### 5.1 Integrated approaches

Finally, our approach involves writing IL measurement features alongside the IRR system, directly integrating measurement taking with system source-code. With the integrated approach, IL can be measured from $t_1$ to $t_5$ (see Figure 2).

When an interaction is registered on the client application of the IRR system, a GUID and stopwatch should be created, and the stopwatch started. Together, the GUID and stopwatch must be stored in a data structure such as a dictionary, where each GUID is a unique key. The GUID should be included in the interaction message sent to the rendering server. The first result message from the server must include the initiating GUID when sent back to the client. On the client, any arriving messages from the server must be inspected for a GIUD. If one exists, the corresponding stopwatch should be extracted from the dictionary and stopped. IL can then be measured as the total number of milliseconds elapsed on the stopwatch.

### 5.4 Summary

Measuring IL is a critical, yet challenging task. We have defined three categories of approach to measuring IL and provided an example of each approach: integrated, observer and hardware.

The integrated approach can only be used when source code is available. An approach such as that used by Chen et. al may be employed, but only when there is certainty that a specific action will result in a corresponding specific on-screen update. The hardware approach can be used to measure IL when source code is not available, but it often involves the access to and setting up of expensive technical equipment. The observer approach can be used regardless of whether or not source code is available, does not require any programming or specific knowledge, and needs no expensive tools and/or hardware, and does not involve setting up of equipment. The following indicates the advantages and disadvantages the three approaches in more detail:

### Integrated

**Advantages**: Useful for debugging system bottle-necks as measurements can be taken at various points in the system;

Doesn't require complex OS input event stream hooking; Provides accurate IL measurements if IDL and DL are not important; No expensive hardware required; Allows for more control and the ability to take measurements at specific system locations.
**Disadvantages**: Requires access to source-code; Will increase bandwidth usage, even if just a little; Will require modification to IRR system and affect it in some way; Ignores IDL and DL. Technical programming knowledge is required.

### Observer
**Advantages**: Can be automated with simulated interactions and automated logging; Might be cheaper than hardware approach as no expensive hardware is required; Provides full event-to-result latency measurements, including DL.
**Disadvantages**: IRR system will be impacted, even if slightly (due to hooking the Graphics API); Difficult to associate an action with a particular frame update; Has measurement resolution equal to the display refresh rate. Ignores IDL.

### Hardware
**Advantages**: Potentially no impact on the IRR system. Provides full event-to-result latency measurements, including IDL and DL.
**Disadvantages**: Not easy to automate; Requires human intervention, which may lead to unreliable measurements; Requires hardware which may be expensive and difficult to set up; Domain-specific knowledge is required to install, configure and use the needed hardware.

## 6 AN INTERACTIVE REMOTE RENDERING SYSTEM

We built an IRR system to provide a platform to test our latency model and to aid in the development of latency simulation, as well as to validate a latency measurement approach introduced later. Our IRR system was built using the popular game engine, Unity3D. We chose Unity3D over the game engine Unreal Engine 4 (UE4) because after comparing the two, we found that Unity3D provides an easy way to access frame data, as well as lower frame-capture (Unity3D: 5.04ms, UE4: 34.92ms) and image compression (Unity3D: 12.69ms, UE4: 59,64ms) times. Comparisons were conducted on our system, which runs Windows 10 in Bootcamp mode, on a 15" MacBook Pro 2017 with a dedicated Radeon Pro 560 (4GB DDR5 RAM) graphics card.

The system is composed of a simple client/server architecture. Both the client and server applications are made with Unity3D. Communications between the client and server applications are performed using TCP. We chose TCP over UDP for its reliability and guarantee that messages sent will be received, and that they will arrive in the order in which they were sent. We measure IL using the integrated approach described in section 5.1 and go into more detail in 6.1.

### 6.1 The client application

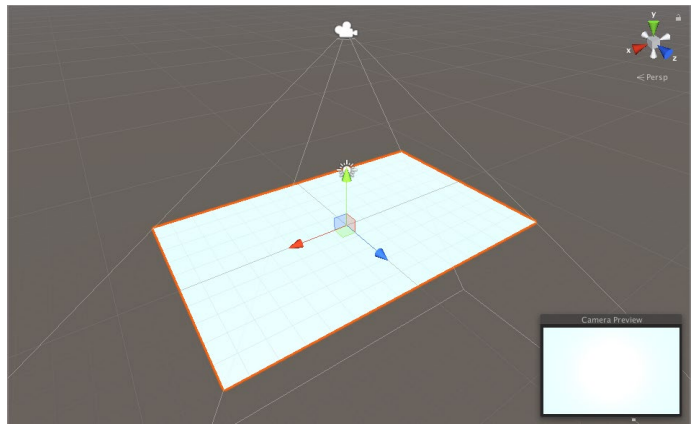

Figure 3. The scene view of our client application. A 2D plane present with a camera positioned above it. The frame data received from the server is converted to a texture and applied to the plane, which the user observes through the camera.

In order to ameliorate cross-platform compatibility issues and to allow us more control, we used Unity3D for the client application too. The client application consists of a simple scene with a camera and a plane (Figure 3). The plane is used to display the images from the server, which are textured mapped onto the plane before being rendered from the camera. There are two modes of interaction: manual (using the keyboard) and automatic (using a template of predefined interactions). In automatic interaction mode, interactions are fed to the system via a template which consist of either an "a" or a "d" per line. The "a" represents "left" and the "d" represents "right" rotation. The interactions are loaded into a queue when the application starts up. At some defined interval, the following procedure takes place:

- Dequeue an interaction
    - Create a "PendingInteraction" object
        - Assign the interaction
        - Create and assign a GUID
        - Create a byte array for image data
        - Create and start a Stopwatch
- Add the PendingInteraction to a dictionary<GUID, PendingInteraction>
- Create a "NetworkMessage" object
    - Assign the interaction
    - Assign the same GUID as above
- Serialize the NetworkMessage
- Send the NetworkMessage to the server.

In the above, a PendingInteraction $PI$ is an interaction performed but which has yet to have a result arrive from the server. The NetworkMessage $M$ passed between the client and server applications has three properties: interaction (string), id (GUID) and frame (byte array). Each $PI$ is transmitted to the server but is also saved to a local Dictionary<GUID, PI>, which we refer to as the Client Interaction Buffer (CIB). When a message arrives on the client, it is deserialized and added to an "arrivedMessages" FIFO queue, referred to as the Client Frame Buffer (CFB).

On each iteration of the Unity3D FixedUpdate (update rate 1ms) method on our client application, the CFB is inspected for any new messages. If the number of elements in the CFB is greater than zero, it is dequeued and the corresponding $PI$ (with matching GUID) is retrieved from the CIB. The stopwatch attached to the $PI$ is then stopped and $PI$ is added to a list called "results", which is later used to calculate IL. The received image data is then applied to the plane and the update scene is presented to the user. Admittedly, the use of Unity3D as a client application does mean that we have to render the scene image as a texture map on a plane before it is presented to the viewer, but we chose to use the same framework for both client and server applications to reduce complexity, to allow for more control, and to gain access to certain Unity3D features such as data types and compression/decompression utilities. Finally, once all pending interactions have been accounted for (all corresponding results have been received and had a measurement logged), the system sends one final message instructing the system to shut down.

## 6.2 The server application

The server application is another Unity3D application, which has a scene consisting of a camera and a large 3D cube. The server application listens for NetworkMessage's arriving from the client and when a message arrives, it is deserialized and an event is raised, signaling to the application that the scene must be updated. To update the scene, the interaction of the arrived message is inspected. If the interaction is an "a", the cube is rotated left. If the interaction is a "d", the cube is rotated right. Once a transformation to the cube is complete, the render result is captured (the game view, shown in Figure 4), compressed to JPG at 75% compression ratio (this is the default supplied by Unity3D), and added to the arrived message. The message is then serialized and finally, sent back to the client.

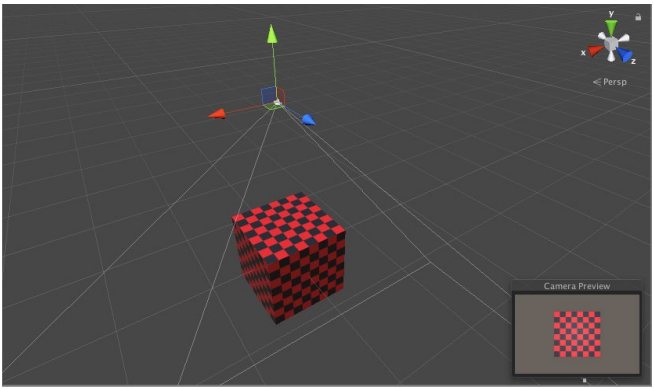

Figure 4. A scene view (top left) of the server application shows a 3D cube with a camera positioned above it. In the game view (top right, bottom left, bottom right) the cube rotates according to interactions received from the client.

## 7 SIMULATING INTERACTION LATENCY

As mentioned earlier, measuring and ensuring minimal IL is a critical task when building IRR systems. However, it is not practical to perform measurements over WAN as there is no control over the amount of latency introduced, it is difficult to repeat experiments and measurements may be influenced by various factors that are outside of our control. Therefore, in order provide a suitable and stable environment for measuring and testing IL, we built a NL simulator with tunable latency parameters.

Simulating interaction latency is not as straightforward as inserting a delay either before or after an interaction has occurred. Even though some authors appear to suggest that this is exactly what they did (for example: [25][11]). Instead, it is critical that interactions are delayed in an asynchronous manner. Figure 5 illustrates how asynchronous processing of interactions results in their inter-arrival times remaining constant. This is important because future interactions must not take longer to complete than previous ones, unless specifically designed to do so. If interactions are delayed synchronously, the amount of delay each interaction experiences will start to increase after the first interaction. However, this can only occur when the delay between interactions (Send Delay, SD) is less than IL. If $SD > IL$, synchronous processing may be used as a backlog of to-be-processed interactions will never occur. Consider the synchronous example of Figure 5.

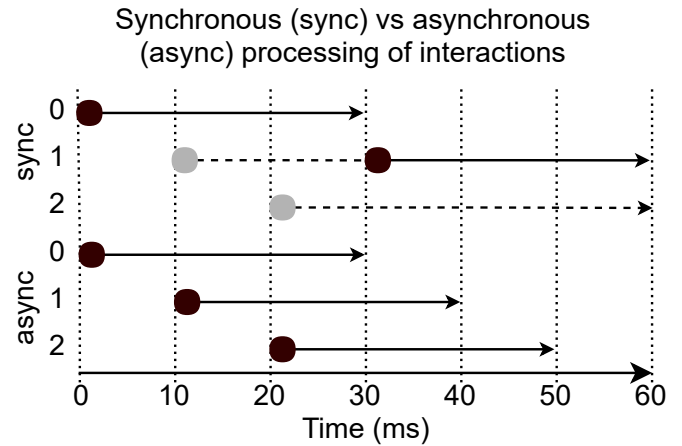

Figure 5. Synchronous vs asynchronous processing of interaction messages. If interactions are processed synchronously and arrive at a rate greater than the total interaction latency, each message will be delayed by an increasing amount as the message buffer starts to form a backlog. With asynchronous processing, the inter-arrival rate of interaction messages is constant.

In the synchronous example, each interaction will be delayed sequentially, thereby adding an additional delay to future interactions and resulting in a backlog. This additional delay is represented by the dashed line in Figure 5. With each interaction talking longer than the previous (due to a growing backlog), we can calculate the expected delay for a given interaction entering the latency simulator with

$$I_{expectedDelay} = \text{NL} + i(\text{NL} - \text{SD}) \qquad (5)$$

where $i$ is the interaction number. Therefore, the delay experienced by each interaction will increase by $NL - SD$, per interaction.

The solution to this issue is to process the incoming messages asynchronously. Unfortunately, asynchronous programming is not easy to implement and can lead to concurrency issues such as interaction $i_{n+1}$ being delayed and processed before interaction $i_n$. The reason for this is that there is no guarantee *when* a task will be started, only that it *will* be started.

To enable us to ensure that interactions will be processed, and their results delivered to the client (from the server) in the order in which they were created and sent, we built a latency simulator which keeps track of the number of received ($M_r$) and processed ($M_p$) messages. Each message that arrives has $M_r$ assigned to its *messageNumber* property. Before a message is released, the latency simulator checks its message number, and it is not released until it is equal to $M_p$ + 1. Both $M_r$ and $M_p$ are initialized to 0.

Our latency simulator has an input *Delay(message, duration)* and an output *MessageReady(result)*, which is an event raised when a message has been delayed. *Delay* creates an object called LatencySimulatorResult $r$ which has two properties: *message* and *messageNumber*. The incoming message is assigned to $r.message$ and $M_r$ to $r.messageNumber$. $M_r$ is then incremented. If $latency = 0$, the *MessageReady(r)* event is immediately fired signaling to any subscribers that a result is ready. If $duration > 0$, a thread is created and $r$ is passed to it. When the thread starts, it immediately sleeps for the specified *duration*. Next, the thread waits until $M_p + 1 = r.messageNumber$. The *MessageReady(r)* event is then raised and finally, $M_p$ is incremented.

The pseudo code for our latency simulator is as follows:

```
Mr, Mp = 0
Delay(message, duration)
    LatencySimulatorResult r
    r.message = message
    r.messageNumber = Mr
    increment Mr
    if duration = 0
        raise event MessageReady(r)
    else Thread(r, duration)
Thread(r, duration)
    sleep(duration)
    while(r.messageNumber ≠ Mp +1)
        sleep(1)
    raise event MessageReady(r)
    increment Mp
```

## 8 RESULTS

So far, we have introduced a model and method for simulating latency, described our IRR system architecture and discussed how to measure IL using the integrated approach. In this section, we present our experimentation results. First, we present the baseline latency of our Unity3D IRR system. We then evaluate our latency simulator, which is integrated into the IRR system, by introducing specific delays and comparing them with real-world measurements. All measurements are performed using the integrated approach described in section 5.1.

### 8.1 IRR Base IL

We first measured the base latency of the system which we described in section 6. We did this using the integrated approach, taking measurements at $t_1$ and $t_6$ (as described in Figure 2). In this first experiment, we ran the client and server applications on our local system (located in Cambridge, UK) by feeding into it 1000 predefined actions using the automatic mode. We chose an arbitrary interaction rate of 10 interaction per second (or 1 interaction every 100ms).

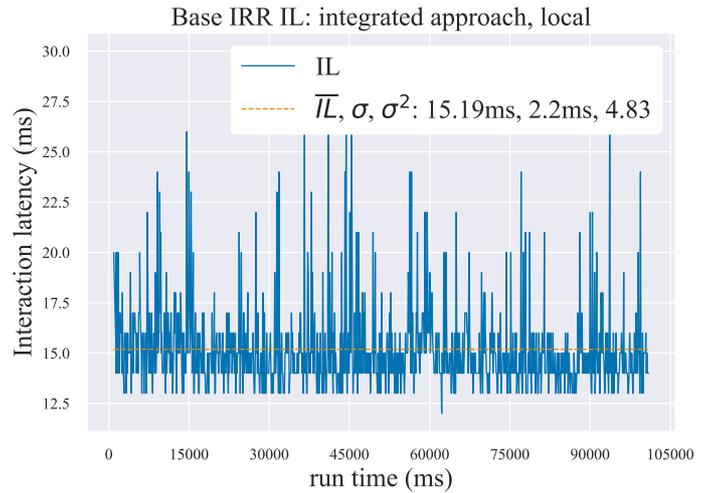

Figure 6. Base interaction latency measured with the integrated approach. Simulation was run on a local machine. Mean IL of 15.19ms, a standard deviation of 2.2ms and a variance of 4.83.

The mean measured base IL ($IL_{base}$) was measured to be 15.19ms. We then performed the same experiment, but over WAN, with the server application on an Amazon EC2 g2.2xlarge instance located in Northern California.

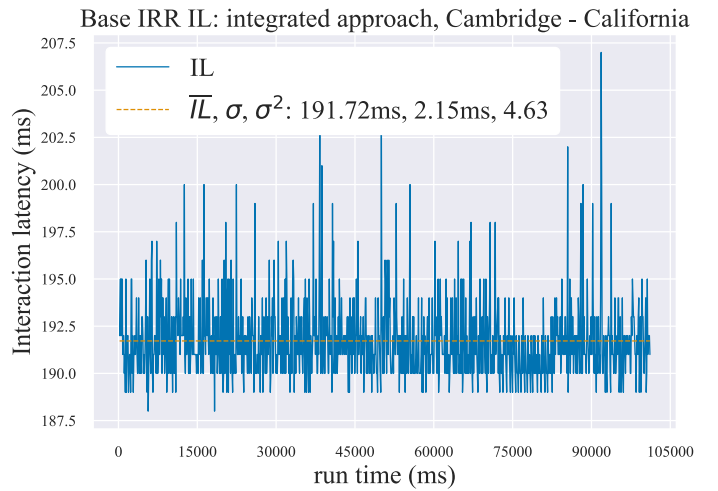

Figure 7. Base IL measured with the integrated approach over WAN between Cambridge, UK and Northern California, USA. The mean network delay (NL) between the two locations is 174ms. We measured a mean IL of 191.72ms with a standard deviation of 2.15 and variance of 4.63.

We ran this experiment three times, one early morning, one in the afternoon and one at night. We did this to counter variation in broadband speeds commonly experienced as "throttling", most often encountered during peak traffic hours. Using the Windows Ping utility (at the same 3 time points mentioned previously), we found that the average NL between Cambridge, UK and the Amazon EC2 server located in Northern California, US is 174ms. The mean IL measured using the integrated approach was 191.72ms. This clearly demonstrates the impact of NL on IL.

### 8.2 Latency simulator

After establishing baseline latencies for the IRR system described in Section 6, we set out to evaluate our latency simulator. To do this, we again used the Unity3D IRR system and performed 3 experiments on a local machine, each with a different latency parameter: 50ms, 100ms and 174ms (the same value as measured between Cambridge and Northern California, using the Ping utility). Our expectation was for our results to show that $\overline{IL} = NL + IL_{base}$.

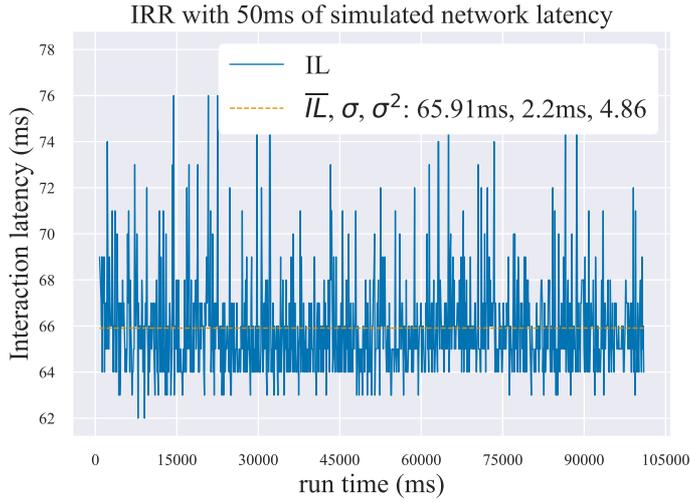

Figure 8. Interaction latency measured using the integrated approach with an additional 50ms simulated latency. Mean measured IL was 65.91ms with a standard deviation of 2.2ms and a variance of 4.86.

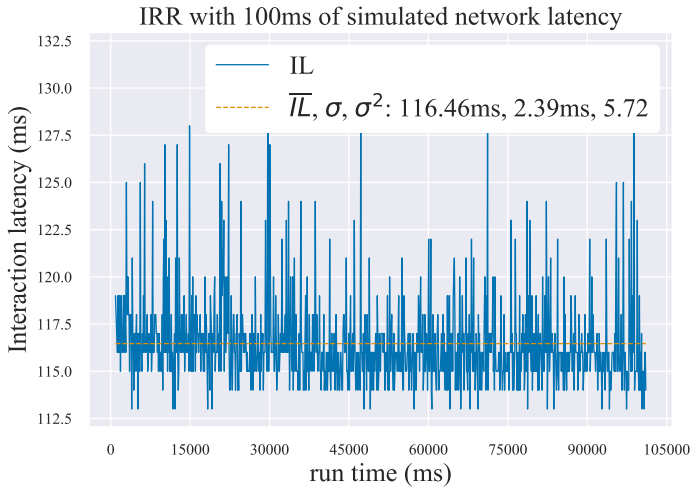

Figure 9. Interaction latency measured using the integrated approach with an additional 100ms simulated latency. Mean measured IL was 116.46ms with a standard deviation of 2.39ms and a variance of 5.72.

In both Figure 8 and Figure 9 we notice that the mean IL measured is ≈ $16ms$ more than the simulated NL, which is an indication that the latency simulator is performing as expected since our measured base latency is ≈ $15ms$. We next compared our IL measurements over WAN (between Cambridge and Northern California) with our simulated IL of 174ms.

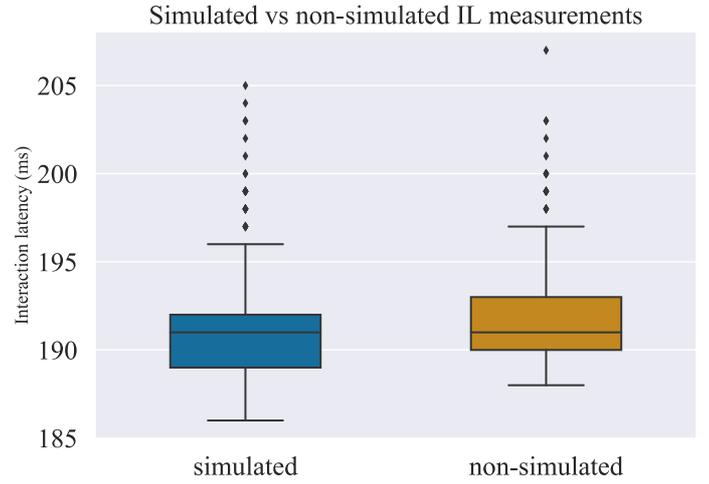

Figure 10 Boxplots showing means of simulated vs non-simulated IL measurements.

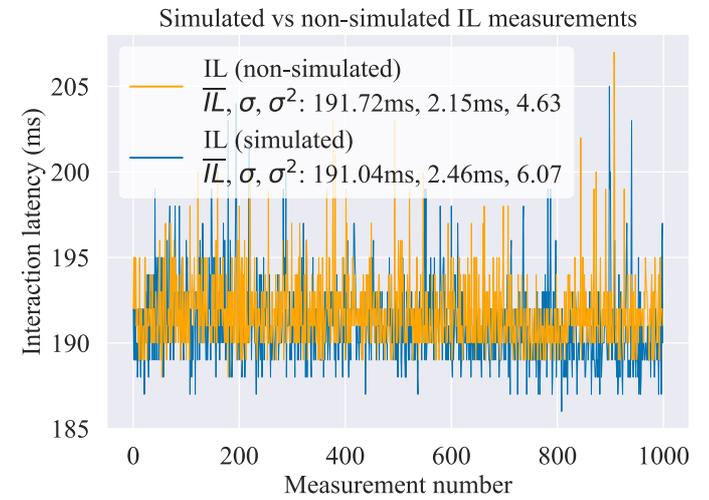

Figure 11. Simulated vs non-simulated (WAN) Interaction latency measured using the integrated approach. Zoomed-in view as signals are nearly identical and overlap otherwise. For the simulated latency, a mean IL of 191.04ms with a standard deviation of 2.46ms and a variance of 6.07 was measured. For the non-simulated latency, a mean IL of 191.72ms with a standard deviation of 2.15ms and a variance of 4.63 was measured.

Figure 10 and Figure 11 shows that the latency simulator produces a delay in the IRR system comparable to that introduced by WAN, with the difference in means of less than 1 millisecond. This is a clear indication that the latency simulator is producing a delay similar to that of a real-world network.

## 9 A SOFTWARE-BASED INTERACTION LATENCY MEASUREMENT TOOL (LMT)

Existing approaches to measuring IL, such as described by Chen et al in [21], are not generic enough for application to a wide variety of IRR systems. For example, when source code is not available, this approach is simply not suitable. Additionally, the integrated approach we used above will affect the IRR system, and therefore in some small way, impact IL. For instance, interactions performed must be tagged with an ID, duplicated and stored (one of each with a stopwatch), thereby increasing the demand on the system resources and increasing the size of the messages transmitted over the network. A rough estimate of the expected IL can be made with:

$$IL \approx RTT + \bar{R} \qquad (6)$$

A rough estimate of IL where $RTT$ is the Round-Trip-Time between the client and the server (measured with Ping) and $\bar{R}$ is the mean rendering time.

Such a rough estimate will not necessarily inform us whether there is more or less latency than expected, nor if there is a need for further investigation. Therefore, our aim is to determine IL without interfering with the IRR system it is measuring and to do so by calculating the time difference between a predefined user input and the moment an image update has been displayed on the monitor/display device, by monitoring a region of interest (ROI) on the screen. To do this we developed a software-based approach to measuring IL, which improves on previous approaches by negating the need to monitor a fixed set of pixels, is generic in that any interaction can be mapped as the input to the IRR system, and can be used in both local and remote desktop scenarios.

Our latency measurement tool (LMT), when initialized, displays a red bounding box (which we refer to as a reticle) of 50x50 pixels (Figure 12 left). On application start-up, the pixels within the box are captured 1000 times, as frequently as possible. This frame-capturing is referred to as the Capture Period ($CP$). From these initial measurements, we calculate the average capture time $\overline{C_t}$ to be 16ms. This is lowest capture time that we can achieve with our setup as our display runs at a maximum of 60Hz. A key is bound such that it has no effect on the IRR system; this key is used to initiate $CP$. We could trigger the $CP$ with an interaction, however the first capture in each $CP$ woud have nothing to be compared with, presenting issue. Therefore, we chose to separate the initiation of the $CP$ and the interaction. The $CP$ executes for the lifespan of the application process and results in a set of captures $CP = \{C_1 \dots C_n\}$ where $C$ is a single capture of 50x50 pixels.

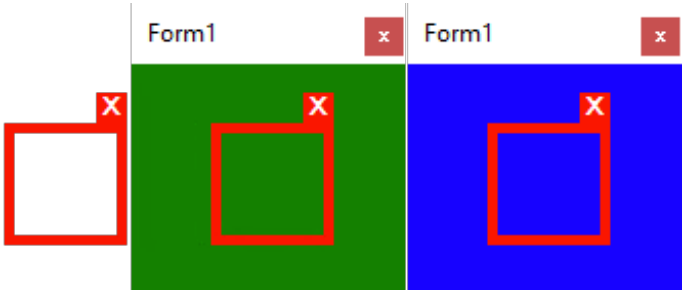

Figure 12. A tool built to measure the baseline performance of our latency measurement tool. The reticle (red bounding box) is used to select pixels for monitoring. The reticle is placed over the green window (center) which changes to blue (right) when an action is performed. The IL is measured as the delay between performing the action and the window changing from green to blue.

After the initial calibration step, the tool hooks into the OS-level keyboard event stream and raises an event when an input is detected. A global stopwatch is also created and started.

We then move the reticle over the green window of a simple Windows application whose background color changes when a key is pressed (Figure 12 center), and manually initiate capturing. Each capture is tagged with a timestamp from the global stopwatch.

While capturing is in progress, a key is pressed (representing an interaction), which causes the background color of the green window to change to blue (Figure 12 right). When an interaction $I$ is detected, the time of the interaction $I_t$ is recorded with the global stopwatch. When measurement taking is complete, the interaction times are matched with capture times. $C_n$ is compared with $C_{n-1}$ on a per-pixel basis. If $C_{n[i,j]} \neq C_{n-1[i,j]}$, where $[i,j]$ represent pixel coordinates, we record IL:

$$IL = C_{n_t} - I_t \qquad (7)$$

Where $C_{n_t}$ is the n<sup>th</sup> capture at time $t$ and $I_t$ is the interaction at time $t$. Since this approach directly compares two frames for differences between them, the reticle must be carefully positioned such that no pixel changes occur unless caused by an interaction. This results in an issue where, for example, background imagery changes without user input (e.g. swaying grass or particles in a simulation): each capture is different and therefore determining which frame results from an interaction is impossible.

To circumvent this problem, we calculate a Peak Signal to Noise Ratio (PSNR) for each captured image. We do this by considering image pairs such that:

$$psnr(C_n) = \begin{cases} 100, & if\ n = 0 \\ PSNR(C_n, C_{n-1}), & otherwise \end{cases} \qquad (8)$$

Where the PSNR is a function which returns a $psnr$ calculated with:

$$PSNR(C_n, C_{n-1}) = \begin{cases} 100, & if\ \epsilon = 0 \\ 20 * \log_{10} \frac{255}{\sqrt[2]{\epsilon}}, & otherwise \end{cases} \qquad (9)$$

Where $\epsilon$ is the Mean Squared Error between captures $C_n$ and $C_{n-1}$.

In a static environment, $C_n$ will have a $PSNR$ of 100 when it is identical to $C_{n-1}$. When the scene updates due to an interaction, the $PSNR$ will drop according to how dissimilar the two captures are. In a non-static environment, no two images will be identical due to small scene variations. However, the difference between two captures which results from an interaction should be detectable in the form of a dramatic and abrupt drop in $PSNR$.

If each capture has an associated $PSNR$ and a timestamp, then IL can be measured as the difference in time between an interaction timestamp and the timestamp associated with the first detected drop in $PSNR$ for a given capture:

$$IL_n = C_t - I_t, when\ |C_{psnr} - C_{psnr-1}| > \theta \qquad (10)$$

Above, $I_t$ is the timestamp of the interaction and $C_t$ is the timestamp of the of the first capture which drops below some threshold $\theta$, which is the mean $PNSR$ at rest, without interaction.

## 9.2 Latency Measurement Tool testing

To evaluate our LMT, we created a simple test application which on keypress, changes its background color from green to blue (we mention this in section 8). There were no other tasks or applications in operation during experimentation and the color change should therefore have been immediate. On a local machine, we placed the reticle (red bounding box) of our LMT over the green background of the test application and performed 50 manual key presses. We use this test application primarily to validate our LMT with the expectation that our measurement results should indicate that $IL \approx \overline{C_t}$.

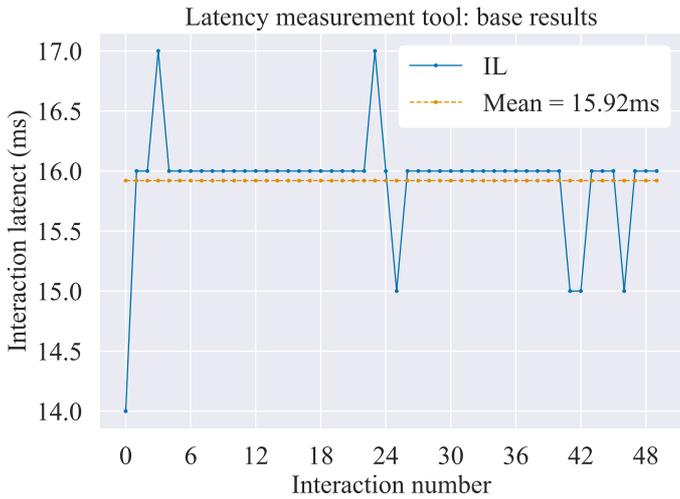

Figure 13. Latency measurement tool base results collected using a test-application which simply changes a Windows Form background color from green to blue, when a key is pressed. We measured a mean IL of 15.92ms.

Indeed, our LMT measured an average IL of 15.92ms, as can be seen in Figure 13, which is very close to the expected $\overline{C_t}$ of 16ms We repeated the experiment and collected 10 new measurements – these interactions were performed manually and were collected for comparison with those of the IRR system in an experiment described later. During the analysis of the results obtained with the LMT, we found clear visual indications that $PSNR$ does indeed drop enough to enable the positive identification of frames resulting from interaction. For example, the following shows $PSNR$ values from data collected for 10 interactions collected over 20 seconds:

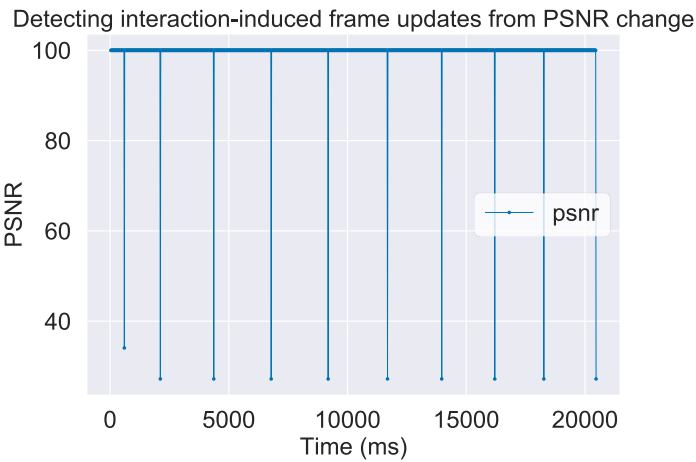

Figure 14. 10 interactions performed over 20 seconds. Clear PSNR drops indicate a frame resulting from an interaction.

In Figure 14, $PSNR$ drops dramatically from 100 to approximately 27 when a frame is significantly different from the one before it (due to interaction). When we zoom into the data, we find that there is a visible delay between when an interaction is performed and when the on-screen image is updated:

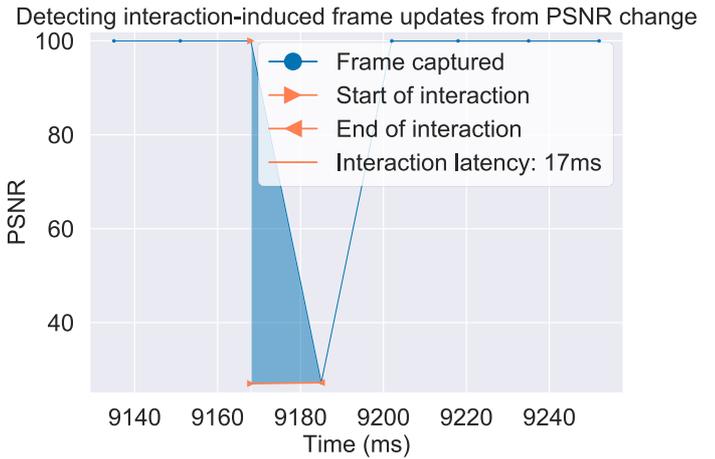

Figure 15 PSNR used to identify change in scene from interaction. Interaction is performed and 17ms later, a drop in PSNR is detected, indicating IL.

Having validated our LMT with test application, we next tested our LMT on the IRR system (described in section 6) with the simulated latencies of 50ms and 100ms. Since our LMT does not simulate interactions and therefore requires input from us, we collected 10 measurements per experiment. During these experiments, we also collected measurements using the integrated approach so that we can compare them with those obtained using the LMT.

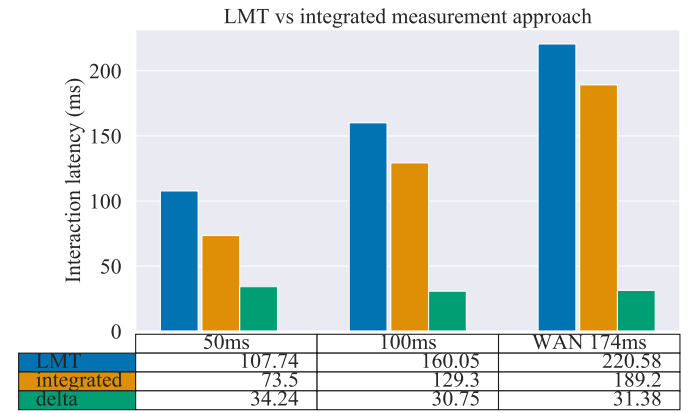

| | 50ms | 100ms | WAN 174ms |
|---|---|---|---|
| LMT | 107.74 | 160.05 | 220.58 |
| integrated | 73.5 | 129.3 | 189.2 |
| delta | 34.24 | 30.75 | 31.38 |

Figure 16. Comparison of measurements between LMT and integrated approach described in Section 5.1. Measurements reported are averages of 10 measurements per experiment. Delta is the difference between the two measurement approaches.

In Figure 16, we compare the results of the LMT with the integrated approach (method described in section 5.1 and expanded on in section 6.1) for simulated latencies of 50ms and 100ms. We then compare the measurements of the obtained with the two approaches by instead of simulating latency, we use a real-world network (WAN), with the server located in Northern California. Both the integrated and LMT measurement pairs were collected from the same experiment, however due to the measurements being collected via different applications, it is not possible to present the results along a unified timeline. From the figure, the LMT is shown to capture more latency than the integrated approach, and the delta (difference in IL measured between the LMT and integrated approach) remains relatively consistent across all latencies, whether simulated or not.

# 10 DISCUSSION

## 10.1 On interaction latency measurement

Latency measurment can be challenging to preform, especially if there is no access to the source code of the application. We have described three approaches to measuring IL, provided an example of each and introduced a novel software tool which expands on and improves the observer approach.

The integrated approach is robust in that the developer has complete control over where measurements are performed and can collect measurements at various points in the pipeline. However, in the case where source code access is not available, the hardware approach is a suitable alternative so long as the correct hardware (high-speed camera, tracker, LED, etc.) is available to the operator and care is taken to set up and perform measurements accurately. The observer approach (such as in the example of Chen; described in section 5.2), is useful but has a number of issues and lacks the ability to consider measurements of applications consisting of moving entities which are not coupled to user interaction (e.g. grass blowing in the wind). We expanded on this approach by capturing screen pixels (by reading from the framebuffer) within a user-controlled reticule and by monitoring PSNR values for the identification of interaction results arriving from the server.

## 10.2 On our IRR system

We measured our IRR system using the integrated approach and while we believe that our measurements are reliable, there is always noise present in IRR systems (e.g. rendering delays) and our system is no exception. For example, in Figure 6 and others, there is a large amount of variability in the data. This suggests that there is an underlying process which is causing instability in our IRR system. The most likely cause is our chosen method for "pausing" a thread. To do this, we use the standard .NET Thread.Sleep() function. Unfortunately, Windows-based machines are not real-time operating systems (OS), providing a thread time-slice of 15ms. As a result, the sleep function has a resolution of 15ms. It has been noted that the time-slice can be configured to 1ms by using timeBeginPeriod and timeEndPeriod, however, we were reluctant to do this as it would affect the thread time-slice OS-wide and may have unexpected consequences. While we believe that this system noise is of little concern, we would like to reproduce and further verify our approach with a real-time OS.

## 10.3 On simulating latency

Our latency simulator was built with the aim of providing researchers and developers with a way to test and evaluate latency compensation techniques, as well as to evaluate our proposed LMT.

We have measured our latency simulator and are confident that it behaves reasonably well when compared with our WAN experiments. In all experiments measured using the integrated approach, the mean IL was $\approx 15ms\ to\ 18ms$ (e.g. Figure 8 and Figure 9) above the introduced latency. This is the expected result as our IRR system has a similar mean baseline. Our experiment using WAN and the integrated approach again confirm that our latency simulator operates well (Figure 11), this time with a mean difference of just $0.68ms$ – a negligible amount considering a system latency of 170ms. It is clear, though, that there is noise in our results, which we suspect is due to the 15ms time-slice limitation of the Windows OS (described in 10.2).

## 10.4 On our Latency Measurement Tool

Our aim in developing the software-based LMT was to enable researchers and developers to measure IL without any hardware or configuration, while limiting the impact of measurement taking on the IRR system. While we have achieved this goal, there are some notable limitations. For example, the LMT we developed has mean capture time ($\overline{C_t}$) dictated by the monitor refresh rate (60Hz), which in our case was $\approx 16ms$. There is therefore a risk that a measurement will be out by at least $\overline{C_t}$ (if pixel change occurs before or after the capture). Nevertheless, our results when comparing the LMT with the integrated approach indicate that our tool does indeed capture IL with reasonable accuracy. Interestingly, the fact that the LMT captures a consistent amount of latency more than the integrated approach suggests that it may be capable of capturing latency which the integrated approach is not.

At present, performing measurements with the LMT is a very manual and slow process, requiring the operator to physically initiate capturing and then perform interactions. However, it should be possible to automate this process by simulating keypresses.

# 11 CONCLUSION

Before we see the wide-spread adoption of cloud enabled data visualization systems, we need to solve the challenges presented by interaction latency. The ability to easily and accurately measure IL is important and therefore, in this paper, we set out to achieve the following:

**A model for interaction latency.**
We introduce a simple end-to-end model of latency within IRR systems, which we apply when measuring interaction latency.

**A method for simulating interaction latency.**
Being able to simulate IL with controllable latency parameters is important for assisting researchers and developers during the design and implementation phase of IRR systems, as well as when investigating potential latency compensation techniques. We have therefore introduced, tested and validated a method for simulating latency. We find that simulating latency in IRR systems is a non-straightforward process and therefore developed a technique which provides fine-grained control over latency within less than 1ms of a real-world signal. Additionally, we find that asynchronous processing of interactions within the latency simulator is a critical factor. To validate our approach, we conducted multiple experiments and compared real-world latency measurements with our simulator-generated latencies.

**A novel framework for measuring interaction latency using a software-based observer approach.**
We define three latency measurement technique categories: integrated, observer, hardware. Additionally, we describe in detail how to apply the integrated approach for measuring IL, which to the best of our knowledge has not been presented in literature. During measurement experimentation, we found that existing approaches are sufficient when access to source code is available. However, all approaches lack generality and are often complicated to configure and use, limiting applicability. We therefore addressed these concerns by developing a novel general-purpose software-based method (LMT) for measuring end-to-end IL. In fact, we believe that our LMT is suitable for measuring the IL of any display system (local or remote) and can even be used on web applications such as YouTube and NetFlix.

With these findings, we hope that simulating and measuring latency is made easier for the developer and research community, and we anticipate applying these introduced methods to our future research where we investigate minimizing latency in IRR systems.

## ACKNOWLEDGEMENTS

This work was supported by the Engineering and Physical Sciences Research Council, Centre for Doctoral Training in Cloud Computing for Big Data [grant number EP/L015358/1]

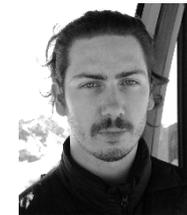

**Richard Cloete** Ph.D student (Newcastle University) Computer Science, B.Sc. (University of Westminster, 2012) 1:1 honours in Software Engineering. After graduating in 2012, he joined a software house in London where he built scientific software for the medical industry. He is currently a EPSRC funded final-year Ph.D student with the Centre for Doctoral Training, Newcastle University at Newcastle University, where he is studying latency in Interactive Remote Rendering systems.

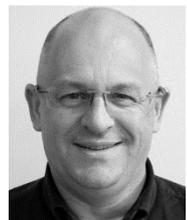

**Nicolas S. Holliman** Ph.D. (University of Leeds 1990) Computer Science, B.Sc. (University of Durham, 1986) joint honours Computing with Electronics. He worked as a software researcher at Lightwork Design Ltd., was Principal Researcher at Sharp Laboratories of Europe Ltd, he was Reader in Computer Science at Durham University and then Professor of Interactive Media at the University of York. He is currently Professor of Visualization at Newcastle University where he heads the Scalable Computing research group and is also a Fellow of the Alan Turing Institute in London. He has published academic articles and patents in visualization, human vision, computer vision, highly parallel computing and autostereoscopic 3D display systems. He is a member of the IS&T, the ACM, a fellow of the Royal Statistical Society and a member of the IEEE Computer Society.